\documentclass[preprintnumbers,nofootinbib,superscriptaddress,10pt]{revtex4}
\usepackage[dvipdfmx]{graphicx} 
\usepackage{amsmath,amssymb,amsfonts, bm, tikz} 

\def\a{\alpha} \def\b{\beta} \def\g{\gamma}  \def\d{\delta} \def\D{\Delta} \def\e{\epsilon}                        

\def\dg{\dagger}  \def\nn{\nonumber}

\usepackage{color}

\begin{document}
%%%%%%%%%%%%%%%%%%%%%%%%%%%%%%%%%%%%%%%%

\title{\large Geometric representation of CP phases $\delta_{\rm PDG}, \delta_{\rm  KM}$ in flavor mixing matrix \\ and its sum rule by alternative unitarity triangle and quadrangle}

\preprint{STUPP-25-287}
%%%%%%%%%%%%%%%%%%%%%%%%%%%%%%%%%%%%%%%%

\author{Masaki J. S. Yang}
\email{mjsyang@mail.saitama-u.ac.jp}
\affiliation{Department of Physics, Saitama University, 
Shimo-okubo, Sakura-ku, Saitama, 338-8570, Japan}
\affiliation{Department of Physics, Graduate School of Engineering Science,
Yokohama National University, Yokohama, 240-8501, Japan}

%%%%%%%%%%%%%%%%%%%%%%%%%%%%%%%%%%%%%%%%

%\date{\today}

%%%%%%%%%%%%%%%%%%%%%%%%%%%%%
\begin{abstract} %%%%%%%%%%%%%%%%%%%%%
%%%%%%%%%%%%%%%%%%%%%%%%%%%%%

In this letter, we present a geometric representation of the CP phases $\delta_{\rm PDG}$ and $\delta_{\rm KM}$  in the PDG and Kobayashi--Maskawa parameterizations of the flavor mixing matrix  in the complex plane.  
The sum rule with the unitarity triangle $\delta_{\rm PDG} + \delta_{\rm KM} = \pi - \alpha + \gamma$ is expressed as a quadrangle, which is a  combination of a unitarity triangle and an alternative triangle. 
Through the unitarity quadrangle, the CP phases are also  identified with specific geometric angles. 
Furthermore, a new set of inverse unitarity triangles is defined from the inversion formula of a unitary matrix $U^{\dagger} = U^{-1}$. 
These novel triangles contain standard angles of the form $\arg [U_{\alpha i } U_{\beta j} U_{\alpha j}^{*} U_{\beta i}^{*}]$ and  new angles $\arg [U_{\alpha i } U_{\beta j} U_{\gamma k} / \det U]$, which directly determine nontrivial arguments of the mixing matrix elements.
%

%%%%%%%%%%%%%%%%%%%%%%%%%%%%%
\end{abstract} %%%%%%%%%%%%%%%%%%%%%%
%%%%%%%%%%%%%%%%%%%%%%%%%%%%%

\maketitle

%%%%%%%%%%
\section{Introduction}
%%%%%%%%%%

Unitarity triangles are fundamental observables describing CP violation in particle physics \cite{Buras:2000dm}.
These triangles have provided a geometric picture of CP violation and have been widely used in comparisons with experimental data.
On the other hand, since the essence of CP violation is  encoded in the CP phase $\delta$ of the mixing matrix $V$, the relationship between unitarity triangles and the CP phase has been extensively discussed \cite{Wu:1994di, Fritzsch:1995nx, Hocker:2006xb, Xing:2009eg, Harrison:2009bz, Antusch:2009hq, Frampton:2010ii, Dueck:2010fa, Frampton:2010uq,Li:2010ae,  Zhou:2011xm,Qin:2011ub,Zhang:2012bk, Zhang:2012ys, Li:2012zxa, He:2013rba, He:2016dco, Xing:2019tsn, Benoit:2022qmt, Kaur:2023ypg,  Harrison:2025rkp}. 
To simplify the treatment of CP violation, different parameterizations from the standard ones have also been invented \cite{Qin:2010hn, He:2008td}. 
However, there are no specific representations to express the CP phase $\delta$ as an angle in the complex plane.

In recent years, analyses of rephasing invariants involving the determinant of the mixing matrix $\det V$ \cite{Yang:2025hex, Yang:2025law, Luo:2025wio} show that the CP phase is written in terms of a combination of matrix elements $V_{ij}$ and $\det V$. 
This type of invariant is a generalization of those defined under the condition $\det V = 1$ 
\cite{Chang:2002yr,Kuo:2005pf,Chiu:2012uc,Chiu:2015ega,Chiu:2017ckv,Kuo:2019psm}. 
These developments open the way to understanding the CP phase as a genuinely geometric quantity. 

In this letter, we represent the sum rule $\d_{\rm PDG} + \d_{\rm KM} = \pi - \a + \g$ \cite{Yang:2025ftl} relating the CP phases $\delta_{\rm PDG}$ and $\delta_{\rm KM}$ in the PDG and Kobayashi--Maskawa (KM) parameterizations to the angles $\alpha$ and $\gamma$ of the unitarity triangle as a unitarity quadangle in the complex plane. 
In addition, using the inversion formula of a unitary matrix, 
we define a new class of ``inverse unitarity triangles''  
whose angles are expressed by $\arg [V_{\a i } V_{\b j} V_{\a j}^{*} V_{\b i}^{*}]$ and $\arg [V_{\a i } V_{\b j} V_{\g k} / \det V]$. 
The new angles $\arg [V_{\a i } V_{\b j} V_{\g k} / \det V]$ directly give the nontrivial phases of the mixing matrix elements in the PDG parametrization. 
 Some of relations derived here coincide with sum rules between the CP phases and angles \cite{Yang:2025vrs, Yang:2025cya}. 
While earlier results are obtained without using the inversion formula, this paper demonstrates that the CP phases and angles are represented geometrically by unitarity relations among lower-order terms of matrix elements.

%%%%%%%%%%
\section{Geometric representation of CP phases $\delta_{\rm PDG}$ and $\delta_{\rm KM}$ by unitarity quadrangle}
%%%%%%%%%%

In this section, we present a method to represent the CP phases of the flavor mixing matrix $\delta_{\rm PDG}$ and $\delta_{\rm KM}$ by combining a unitarity triangle with an alternative triangle.
The flavor mixing matrix has nine Euler-angle-like parameterizations \cite{Fritzsch:1997st},
which have matrix elements with trivial phases $V_{\alpha j}, V_{\beta i} \in \mathbb{R}$ 
for a row $\alpha$ and a column $i$. 
Denoting the CP phases in these parameterizations as $\delta^{(\alpha i)}$, 
the rephasing invariant formulae for $\delta^{(\alpha i)}$ are given by \cite{Yang:2025hex, Yang:2025vrs}
\begin{align}
\d^{(\a i )} %& =  (\g_{L1} + \g_{L2} + \g_{L3} - \g_{R1} - \g_{R2} - \g_{R3}) - \arg  \det V \nn \\
%& = \arg [V_{\a 1} V_{\a 2} V_{\a 3} V_{1i} V_{2i} V_{3i} / V_{\a i}^{3}] - \arg  \det V
= \arg \left[ { V_{\a 1} V_{\a 2} V_{\a 3} V_{1i} V_{2i} V_{3i}  \over V_{\a i }^{3} \det V } \right] . 
\end{align}
In practice, the two $V_{\alpha i}$ in the numerator and denominator cancel out, making these phases fifth-order invariants involving $\det V$.
Important examples are the phase of the standard PDG parameterization $\delta_{\rm PDG}$ and of the KM parameterization $\delta_{\rm KM}$, 
\begin{align}
\d^{(11)} = \pi - \d_{\rm KM} = \arg \left[ { V_{12} V_{13}  V_{21} V_{31}  \over V_{11} \det V } \right]   , ~~ 
\d^{(13)} = \d_{\rm PDG} = \arg \left[ { V_{11} V_{12}  V_{23} V_{33}  \over V_{13} \det V} \right]  .  
\end{align}
This method applies not only to the quark mixing but also to the lepton mixing, 
because changing the values of Majorana phases corresponds to rephasings of the neutrino fields, 
whereas the CP phases are written by rephasing invariants. 

With the angles of the unitarity triangles, 
\begin{align}
\a = \arg \left [ - { V_{31}^{} V_{33}^{*} \over V_{11}^{} V_{13}^{*} } \right ]  , ~~
\b = \arg \left [ - { V_{21}^{} V_{23}^{*} \over V_{31}^{} V_{33}^{*} }  \right ]  , ~~
\g = \arg \left [ - { V_{11}^{} V_{13}^{*} \over V_{21}^{} V_{23}^{*} }  \right ] , 
\end{align}
these phases and angles are related by a sum rule  \cite{Yang:2025ftl, Yang:2025vrs}
\begin{align}
\d_{\rm PDG} + \d_{\rm KM} = \pi - \a + \g  \, .
\end{align}
Such a sum rule can also be understood  by simple products and ratios with third-order invariants  \cite{Yang:2025cya}; 
\begin{align}
 \arg \left[- \frac{  V_{12} V_{23} V_{31} }{  \det V } \right]  
& = 
\arg \left[ { V_{11} V_{12}  V_{23} V_{33}  \over V_{13} \det V} \right]  
+ 
\arg \left[ - { V_{13}  V_{31} \over V_{11} V_{33} } \right]
= 
 \arg \left[ { V_{12} V_{13}  V_{21} V_{31}  \over V_{11} \det V } \right]  
+ 
 \arg \left[ -  {V_{23} V_{11} \over V_{21} V_{13}} \right] \nn \\ 
 & %= \chi_{2} + \pi %& = \d^{(13)}  + \Phi_{22} = \d^{(11)} +  \Phi_{32}  
= \delta_{\rm PDG} + \a = \pi - \d_{\rm KM} +  \g \, . 
\end{align}

The form of the sum rule suggests that it can be understood geometrically by unitarity triangles or their combinations. 
Indeed, the relation is interpreted as a quadrangle in the complex plane.
We focus on a unitarity triangle corresponding to the first and third rows, which includes the angle $\alpha$, 
\begin{align}
V_{11}^* V_{31} + V_{12}^* V_{32} + V_{13}^* V_{33} = 0 \, . 
%~~ V_{12}^* V_{32}  = -  V_{11}^* V_{31}   - V_{13}^* V_{33}  \, . 
\end{align}
Considering an element $V_{32}$ as the inverse of the conjugate matrix $V^{\dg}$,
\begin{align}
V_{ij} =  {1\over \det V^{*} }\sum_{k,l,m,n} {1 \over 2 } \e_{ikl} \e_{jmn} V_{km}^{*} V_{ln}^{*}  \, ,  ~~ 
V_{32} =  { V_{13}^* V_{21}^* - V_{11}^* V_{23}^* \over \det V^{*} }  \, ,  
\end{align}
and substituting $V_{32}$ to the unitarity condition,  one obtains
\begin{align}
% - & V_{11}^* V_{31}   - V_{13}^* V_{33} = V_{12}^* V_{32} =     {V_{12}^{*}  V_{13}^* V_{21}^* - V_{12}^{*}  V_{11}^* V_{23}^* \over \det V^{*} }  \, , \nn \\
 V_{11}^* V_{31}  + V_{13}^* V_{33}  - { V_{12}^{*}  V_{11}^* V_{23}^* \over \det V^{*} } +  {V_{12}^{*}  V_{13}^* V_{21}^* \over \det V^{*} } 
 = 0  \, . 
 \end{align}
This procedure corresponds to combining two triangles in the complex plane to form a quadrangle,
 illustrated in Fig.~1. 
  Each angles are found to be
\begin{align}
& \arg \left [ - {V_{11}^* V_{31} \over V_{13}^* V_{33}  } \right ] =  \a \, , ~
\arg \left [  { V_{13}^* V_{33} \det V^{*} \over V_{12}^{*}  V_{11}^* V_{23}^* } \right ] = \d_{\rm PDG} \, , ~ \nn \\
& \arg \left [  {  V_{11}^* V_{23}^* \over  V_{13}^* V_{21}^* } \right ] = \pi - \g \, , ~
\arg \left [ - { V_{12}^{*}   V_{13}^* V_{21}^* \over \det  V^{*} V_{11}^* V_{31} } \right ] = \d_{\rm KM }\, .  ~
\end{align}
The sum of the interior angles of the quadrangle is 
\begin{align}
\a + \d_{\rm PDG} + \pi - \g + \d_{\rm KM} = 2 \pi \, , 
\end{align}
and the sum rule also holds geometrically. 
Such a unitarity relation and the sum rule for the angles can also be verified numerically.

\begin{figure}[t]
\begin{center}
 \includegraphics[width=12cm]{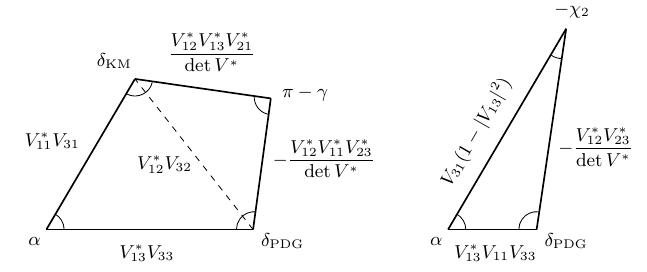}
\caption{Geometric understanding of the CP phases  and the sum rule by a unitarity quadrangle and a new unitarity triangle. 
This figure is drawn without assuming any specific mixings of quarks or leptons, and 
the sum of vectors is defined in the counterclockwise direction. }
\end{center}
\end{figure}

Moreover, by extending two sides of the quadrangle, one can construct a new triangle whose angles include $\delta_{\rm PDG}$ and $\alpha$. 
%
%Using the similarity relations of triangles, one can find the corresponding equations, and
 Here we present the results after simplification.
By substituting 
$V_{22}^{*} = (V_{11} V_{33} - V_{13} V_{31}) / \det V$ 
into the element $V_{31}$ of the inversion formula, 
\begin{align}
V_{31} &= {V_{12}^{*} V_{23}^{*} - V_{13}^{*} V_{22}^{* } \over  \det V^{*} } 
= {V_{12}^{*} V_{23}^{*} \over  \det V^{*} } - V_{13}^{*} (V_{11} V_{33} - V_{13} V_{31} )  \, . 
\end{align}
Since the terms proportional to $V_{31}$ is combined into a single term, 
\begin{align}
 V_{31}  (1- |V_{13}|^{2} ) + V_{13}^{*} V_{11} V_{33}   - {V_{12}^{*} V_{23}^{*} \over \det V^{*}}= 0 \, .
\end{align}
This triangle is also shown in the right panel of Fig.~1. 
The three angles are, 
\begin{align}
\arg \left[  {V_{13}^{*} V_{11} V_{33} \over V_{12}^{*} V_{23}^{*} \det V }  \right ] = \d_{\rm PDG} \, , ~ 
\arg \left[  { V_{12}^{*} V_{23}^{*} \over V_{31} \det V^{*}}  \right ] \equiv - \chi_{2} \, ,  ~ 
\arg \left[  - { V_{31} \over V_{13}^{*} V_{11} V_{33} }  \right ] = \a  \, . 
\end{align}
Therefore, this is a novel unitarity triangle that directly generates $\d_{\rm PDG}$ 
with the angle $\chi_{2}$. 
The relation $\d_{\rm PDG} - \chi_{2} + \a = \pi$ is algebraically trivial and a part of another sum rules~(\ref{sumrule2}).

%%%%%%%%%%%%%%%
\subsection*{Inverse unitarity triangles}
%%%%%%%%%%%%%%%

The inversion formula implicitly contains information on the CP phase, 
because they are characterized by invariants involving the determinant.
From this insight and previous discussions, the {\it inverse} unitarity triangles are defined as follows.
The inversion formula of a unitary matrix $U^{\dagger} = U^{-1}$, 
or an equivalent form $U = (U^{\dagger})^{-1}$ is
\begin{align}
\begin{pmatrix}
 U_{11} & U_{12} & U_{13} \\
 U_{21} & U_{22} & U_{23} \\
 U_{31} & U_{32} & U_{33} \\
\end{pmatrix}
= {1\over \det U^{*}}
\begin{pmatrix}
 U_{22}^* U_{33}^*-U_{23}^* U_{32}^* & U_{23}^* U_{31}^*-U_{21}^* U_{33}^* & U_{21}^* U_{32}^*-U_{22}^* U_{31}^* \\
 U_{13}^* U_{32}^*-U_{12}^* U_{33}^* & U_{11}^* U_{33}^*-U_{13}^* U_{31}^* & U_{12}^* U_{31}^*-U_{11}^* U_{32}^* \\
 U_{12}^* U_{23}^*-U_{13}^* U_{22}^* & U_{13}^* U_{21}^*-U_{11}^* U_{23}^* & U_{11}^* U_{22}^*-U_{12}^* U_{21}^* \\
\end{pmatrix} \, . 
\end{align}
Since the formula implies that the sum of three complex numbers equals zero, 
nine novel triangles are defined in the complex plane as
\begin{align}
\det U^{*}
\begin{pmatrix}
 U_{11} & U_{12} & U_{13} \\
 U_{21} & U_{22} & U_{23} \\
 U_{31} & U_{32} & U_{33} \\
\end{pmatrix}
-
\begin{pmatrix}
 U_{22}^* U_{33}^*& U_{23}^* U_{31}^* & U_{21}^* U_{32}^* \\
 U_{13}^* U_{32}^* & U_{11}^* U_{33}^* & U_{12}^* U_{31}^*\\
 U_{12}^* U_{23}^* & U_{13}^* U_{21}^* & U_{11}^* U_{22}^* \\
\end{pmatrix} 
+ 
\begin{pmatrix}
U_{23}^* U_{32}^* &U_{21}^* U_{33}^* &U_{22}^* U_{31}^* \\
U_{12}^* U_{33}^* &U_{13}^* U_{31}^* &U_{11}^* U_{32}^* \\
U_{13}^* U_{22}^* &U_{11}^* U_{23}^* & U_{12}^* U_{21}^* \\
\end{pmatrix}  = O \, ,
\label{O}
\end{align}
where $O$ denotes the zero matrix.
Since the inversion formula expresses the orthonormality by cross products, 
these triangles are also interpreted as unitarity triangles of cross product, in contrast to the usual unitarity triangles of dot product.

The novel triangles constructed from this matrix equation play the role of the Rosetta stone between the standard unitarity triangles and  CP phases in the mixing matrix. 
By taking the argument of the ratio of the second and third matrix in Eq.~(\ref{O}), 
angles of the usual unitarity triangles are immediately reproduced as 
\begin{align}
& 
\arg 
\begin{pmatrix}
 U_{22}^* U_{33}^*& U_{23}^* U_{31}^* & U_{21}^* U_{32}^* \\
 U_{13}^* U_{32}^* & U_{11}^* U_{33}^* & U_{12}^* U_{31}^*\\
 U_{12}^* U_{23}^* & U_{13}^* U_{21}^* & U_{11}^* U_{22}^* \\
\end{pmatrix} 
- \arg 
\begin{pmatrix}
U_{23}^* U_{32}^* &U_{21}^* U_{33}^* &U_{22}^* U_{31}^* \\
U_{12}^* U_{33}^* &U_{13}^* U_{31}^* &U_{11}^* U_{32}^* \\
U_{13}^* U_{22}^* &U_{11}^* U_{23}^* & U_{12}^* U_{21}^* \\
\end{pmatrix} \nn
\\ & = 
\arg 
\begin{pmatrix}
- U_{22}^* U_{33}^* U_{23} U_{32} &- U_{23}^* U_{31}^* U_{21} U_{33}&- U_{21}^* U_{32}^* U_{22} U_{31} \\
- U_{13}^* U_{32}^* U_{12} U_{33} &-U_{11}^* U_{33}^* U_{13} U_{31}&- U_{12}^* U_{31}^* U_{11} U_{32} \\
- U_{12}^* U_{23}^* U_{13} U_{22} &- U_{13}^* U_{21}^* U_{11} U_{23}&- U_{11}^* U_{22}^* U_{12} U_{21}\\
\end{pmatrix}
+  
\begin{pmatrix}
\pi & \pi & \pi \\ 
\pi & \pi & \pi \\
\pi & \pi & \pi \\
\end{pmatrix}
\equiv \Phi + \Pi  \, . 
\end{align}
Here, $\arg$ is taken for each matrix element individually. 
The matrix $\Phi$ represents the nine angles of the unitarity triangles defined in Harrison et al. \cite{Harrison:2009bz}.

Next, the phase differences between the third and the first matrix are
\begin{align}
& 
\arg 
\begin{pmatrix}
U_{23}^* U_{32}^* & U_{21}^* U_{33}^* & U_{22}^* U_{31}^* \\
U_{12}^* U_{33}^* & U_{13}^* U_{31}^* & U_{11}^* U_{32}^* \\
U_{13}^* U_{22}^* & U_{11}^* U_{23}^* & U_{12}^* U_{21}^* \\
\end{pmatrix} 
- \arg \det U^{*}
 \begin{pmatrix}
- U_{11} & - U_{12} & - U_{13} \\
- U_{21} & - U_{22} & - U_{23} \\
- U_{31} & - U_{32} & - U_{33} \\
\end{pmatrix} \nn
\\ & = 
\arg {1\over  \det U^{*}}
\begin{pmatrix}
- U_{23}^* U_{32}^* U_{11}^* &- U_{21}^* U_{33}^* U_{12}^* & - U_{22}^* U_{31}^* U_{13}^{*} \\
- U_{12}^* U_{33}^* U_{21}^*& - U_{13}^* U_{31}^* U_{22}^* & - U_{11}^* U_{32}^* U_{23}^{*} \\
- U_{13}^* U_{22}^* U_{31}^* &-  U_{11}^* U_{23}^* U_{32}^* & - U_{12}^* U_{21}^* U_{33}^{* }\\
\end{pmatrix}  
\equiv  -
\begin{pmatrix}
\psi_{1} & \psi_{2} & \psi_{3} \\
\psi_{2} & \psi_{3} & \psi_{1} \\
\psi_{3} & \psi_{1} & \psi_{2}  \\
\end{pmatrix} \equiv - \Psi \, . 
\end{align}
We define the phase of the odd permutation including $U_{1i}$ multiplied by $(- 1 / \det U)$ as $\psi_{i}$, and the matrix composed of these phases as $\Psi$ \cite{Yang:2025cya}. 
The negative sign arises from complex conjugation.

Finally, 
\begin{align}
& 
\arg  \det U^{*}
\begin{pmatrix}
 U_{11} & U_{12} & U_{13} \\
 U_{21} & U_{22} & U_{23} \\
 U_{31} & U_{32} & U_{33} \\
\end{pmatrix}
- 
\arg 
\begin{pmatrix}
U_{22}^* U_{33}^* &U_{23}^* U_{31}^* &U_{21}^* U_{32}^* \\
U_{13}^* U_{32}^* &U_{11}^* U_{33}^* &U_{12}^* U_{31}^*\\
U_{12}^* U_{23}^* &U_{13}^* U_{21}^* &U_{11}^* U_{22}^* \\
\end{pmatrix} \nn
\\ & = 
\arg {1\over \det U}
\begin{pmatrix}
U_{11}  U_{22} U_{33} & U_{12} U_{23} U_{31}& U_{13}U_{21} U_{32}  \\
U_{21}  U_{13} U_{32} & U_{22} U_{11} U_{33}& U_{23} U_{12} U_{31}\\
U_{31}  U_{12} U_{23} & U_{32} U_{13} U_{21}& U_{33}U_{11} U_{22} \\
\end{pmatrix} 
\equiv  
\begin{pmatrix}
\chi_{1} & \chi_{2} & \chi_{3} \\
\chi_{3} & \chi_{1} & \chi_{2} \\
\chi_{2} & \chi_{3} & \chi_{1} \\
\end{pmatrix}
\equiv  {\rm X} \, , 
\end{align}
where $\chi_{i}$ denotes the phase of the even permutation including $U_{1i}$ multiplied by $1 / \det U$, and the matrix composed of these phases as X.
Since the sum of interior angles of a triangle is $\pi$,  these three matrices satisfy 
\begin{align}
\Phi  + \Pi  - \Psi + {\rm X} = \Pi \, , ~~ \Phi  = \Psi - {\rm X} \, . 
\end{align}
This matrix identity is confirmed in the previous paper \cite{Yang:2025cya}, by simple sums and differences of arguments without employing the inversion formula. 
It indicates that the angles of standard unitarity triangles $\Phi$ are decomposed in terms of the arguments of the third-order invariants. 

These third-order invariants correspond to the nontrivial arguments of the matrix elements 
$ U^{0}_{\alpha i}$ in the PDG parametrization; 
\begin{align}
\chi_{1} = \arg U_{22}^{0} \, , ~~ \chi_{2} = \arg U_{31}^{0} \, , ~&~  \chi_{3} = \arg [U_{13}^{0}  U_{21}^{0}  U_{32}^{0} ]\, , ~~ \nn \\
\psi_{1} = \arg U_{32}^{0} + \pi \, , ~~ \psi_{2} =  \arg U_{21}^{0} + \pi \, , ~&~  \psi_{3} = \arg [U_{13}^{0}  U_{22}^{0} U_{31}^{0} ]+ \pi  \, . 
\end{align}
Therefore, the angles of the inverse unitarity triangles directly determine the arguments of the mixing matrix. 
In particular, since the CP phase $\d_{\rm PDG}$ is given by
\begin{align}
\d_{\rm PDG} = - \arg U_{13}^{0} = \psi_{1} + \psi_{2} - \chi_{3} \, , 
\end{align}
the phase is determined directly from the angles, rather than from the area of the triangle. 

All nine CP phases are also expressed  by these third-order invariants as 
\begin{align}
\D T  = \Psi + T \Phi \, ,  ~ \D T^{T} =  \Psi + T^{T} \Phi \, ,  
\label{sumrule2}
\end{align}
where the phase matrix $\D$ and the transfer matrix $T$ are
\begin{align}
\D = 
\begin{pmatrix}
\d^{(11)} & \d^{(12)} & \d^{(13)} \\
\d^{(21)} & \d^{(22)} & \d^{(23)} \\
\d^{(31)} & \d^{(32)} & \d^{(33)} 
\end{pmatrix} , ~
T = 
\begin{pmatrix}
 0 & 0 & 1 \\
 1 & 0 & 0 \\
 0 & 1 & 0 \\
\end{pmatrix} . 
\end{align}
Since the phases are also generated from the sum of $\psi_{i}$ and $\chi_{i}$,  the angles of new unitarity triangles contain more information than the standard angles. 

%%%%%%%%%%%%%%
\section{Summary}
%%%%%%%%%%%%%%

In this letter, we present a geometric representation of the CP phases $\delta_{\rm PDG}$ and $\delta_{\rm KM}$ in the PDG and Kobayashi--Maskawa parameterizations of the flavor mixing matrix in the complex plane.  
The sum rule with the unitarity triangle $\delta_{\rm PDG} + \delta_{\rm KM} = \pi - \alpha + \gamma$ is expressed as a quadrangle, a combination of a unitarity triangle and an alternative triangle. 
Each side of the quadrangle is given by a product of matrix elements $V_{ij}$ and $\det V$, and  the sum rule is  visually understood by the sum of four interior angles $\alpha + \delta_{\rm PDG} + (\pi-\gamma) +  \delta_{\rm KM} = 2\pi$. 

Furthermore, a new set of inverse unitarity triangles is defined from the inversion formula of a unitary matrix $U^{\dagger} = U^{-1}$. 
These nine triangles with $\det U$ contain the standard angles  $\arg [U_{\alpha  i } U_{\beta j} U_{\alpha j}^{*} U_{\beta i}^{*}]$ and new angles $\arg [U_{\alpha i } U_{\beta j} U_{\gamma k} / \det U]$, that directly determine the nontrivial arguments of the mixing matrix.

Through the novel unitarity triangle and quadrangle, the CP phases are generated from unitarity relations involving lower-order products of mixing matrix elements, and identified with specific geometric angles. These results provide a direct geometric interpretation for the CP phases and extend the unitarity triangle picture to quadrangles.
It offers a new visual tool to deepen the understanding of CP violation, and such quadrangles are naturally expected for more generalized sum rules.

%%%%%%%%%%%%%%
\section*{Acknowledgment}
%%%%%%%%%%%%%%

The study is partly supported by the MEXT Leading Initiative for Excellent Young Researchers Grant Number JP2023L0013.

%\bibliographystyle{bib/h-physrev50}
%\bibliography{bib/fourzero,bib/onezero,bib/refsym,bib/mutausym,bib/PSGUT,bib/StrongCP,bib/LR,bib/GCP,bib/U(2),bib/flaxion,bib/minimal-natural,bib/chiral, bib/T2HK,bib/CKM2MNS,bib/KMCPV,bib/Kuo}

\end{document}